# Creation and confirmation of Hopfions in magnetic multilayer systems

Noah Kent[1,2]*, Neal Reynolds[1,3], David Raftrey[1,2], Ian T.G. Campbell[1,3], Selven Virasawmy[4], Scott Dhuey[4], Rajesh V. Chopdekar[5], Aurelio Hierro-Rodriguez[6], Andrea Sorrentino[7], Eva Pereiro[7], Salvador Ferrer[7], Frances Hellman[1,3], Paul Sutcliffe[8], Peter Fischer[1,2]*

[1]*Materials Sciences Division, Lawrence Berkeley National Laboratory, Berkeley, CA 94720, USA*

[2]*Physics Department, UC Santa Cruz, Santa Cruz CA 95064, USA*

[3]*Department of Physics, University of California, Berkeley, Berkeley, CA 94720, USA*

[4]*The Molecular Foundry, Lawrence Berkeley National Laboratory, Berkeley, CA 94720, USA*

[5]*Advanced Light Source, Lawrence Berkeley National Laboratory, Berkeley, CA 94720, USA*

[6]*Department of Physics, University of Oviedo, 33007 Oviedo, Spain*

[7]*ALBA Synchrotron, 08290 Cerdanyola del Vallès, Spain*

[8]*Department of Mathematical Sciences, Durham University, Durham DH1 3LE, UK*

*Corresponding authors: Nakent@ucsc.edu, PJFischer@lbl.gov*

**Abstract**

Topological solitons have been studied for decades in classical field theories[1], and have started recently to impact condensed matter physics[2-4]. Among those solitons, magnetic skyrmions are two-dimensional particle-like objects with a continuous winding of the magnetization[5,6], and magnetic Hopfions are three-dimensional topological solitons that can be formed from a closed loop of a twisted skyrmion string[3,7-9]. Whereas intense research is underway with magnetic skyrmions towards a fundamental understanding and potential applications in advanced storage and logic devices[10-15], the experimental creation and confirmation of magnetic Hopfions has been elusive so far. Theoretical models suggest that Hopfions can be stabilized in frustrated or chiral magnetic systems, and that target skymions can be transformed into Hopfions by adapting their perpendicular magnetic anisotropy[3,4].

Here, we present experimental evidence of magnetic Hopfions that were created in magnetic Ir/Co/Pt multilayers shaped into nanoscale disks, which are known to host target skyrmions[16]. The three-dimensional spin texture, which distinguishes magnetic Hopfions from target skyrmions was confirmed by combining two advanced element-specific magnetic X-ray microscopy techniques with about 20-30nm lateral resolution, using X-ray magnetic circular dichroism effect as magnetic contrast mechanism in surface-sensitive X-ray photoemission electron microscopy and bulk-sensitive soft x-ray transmission microscopy[17].

We anticipate that these results will stimulate further investigations of Hopfions with different topologies and their potential application in three-dimensional spintronics devices.

**One sentence summary**

Magnetic Hopfions were created by adapting the perpendicular magnetic anisotropy in Ir/Co/Pt multilayered systems, and their characteristic three-dimensional spin texture was confirmed by combining surface sensitive X-ray photoemission electron microscopy with bulk-sensitive soft x-ray transmission microscopy.

**Main text**

In classical field theories, topological solitons describe stable, particle-like objects that have a finite mass and a smooth structure[1]. Whereas first approaches focused on problems in high-energy physics, recently this concept has seen significant attention in condensed matter research, specifically in ferromagnetic[11,18] or ferroelectric materials[19]. There, the smoothly varying vector field is the magnetization or the polarization of the material, respectively. The topological nature of the field is described by a topological charge which describes the continuous winding of the vector field, i.e. the magnetization in a magnetic topological soliton.

Prominent examples are magnetic skyrmions, which are two-dimensional topologically protected spin textures, where the topological charge, the skyrmion number, is defined as $N_{sk} = \frac{1}{4\pi} \iint d^2 r \mathbf{m} \cdot (\frac{\partial \mathbf{m}}{\partial x} \times \frac{\partial \mathbf{m}}{\partial y}))$ with **m** being the unit vector of the magnetization. The spin textures in ferromagnets are the result of competing interactions, such as exchange and magnetic anisotropy, where the former prefers a parallel alignment of neighboring spins, and the latter favors certain orientation, e.g. given by the crystallography of the materials. Mathematically, in most condensed matter systems, localized structures, such as magnetic skyrmions, would rapidly collapse into linear singularities (Hobart-Derrick theorem)[20]. However, in magnetic materials with broken inversion symmetry, which can be achieved, e.g. through a chirality in the crystallographic structure or at interfaces in low-dimensionality thin films, an additional asymmetric exchange interaction, the so-called Dzyaloshinksi-Moriya interaction (DMI) enables the stabilization of localized magnetic states such as skyrmions with finite sizes. The spin texture in magnetic skyrmions can be described by a continuous rotation of the magnetization in the radial direction from the core at the center pointing antiparallel to the orientation far away from the center. The stability of magnetic skyrmions and their high mobility enabling them to be driven through a material with relatively low current densities makes them promising candidates for future high-density, high-speed, and low-power spintronics applications. Although for most of the current research on magnetic skyrmions, they are considered to be two-dimensional topological solitons, in real systems they are nanoscale cylinders, and their two-dimensional treatment assumes a rigid three-dimensional structure. Other three-dimensional spin textures, such as chiral bobbers[21,22] or skyrmion strings[23], which have recently been found in spin systems have a robust and flexible three-dimensional structure exhibiting, e.g. non-trivial dynamic responses[24].

A generalization of magnetic skyrmions into the third dimension leads to more complex and diverse topological solitons, including rings, knots and links. Hopf solitons, or Hopfions (Fig 1b bottom) are such three-dimensional topological solutions, particularly knot-solitons, that can be classified by another topological invariant, the Hopf number ($Q_H$), which in real space can be expressed as $Q_H = -\int \boldsymbol{B} \cdot \boldsymbol{A} d^3 r$ with $\boldsymbol{B}$ being the emergent magnetic field from the spin texture and $\boldsymbol{A}$ the magnetic vector potential[25]. The Hopf number distinguishes the topology in different Hopfions and can be geometrically interpreted as the linking number of any two closed-loop regions in 3D real space that contains spins pointing in the same direction[2,3]. Consequently, a Hopfion can be viewed as a closed loop of a twisted skyrmion string[26]. Whereas stable Hopfions with sizes in the micrometer regime have been experimentally observed e.g., chiral ferromagnets and liquid crystals[7], the realization of stable Hopfions in chiral magnets or magnetic nanostructures, has been elusive so far, but given the enormous progress in understanding the static and dynamics of magnetic skyrmions in low dimensional magnetic systems and their exploration towards novel spintronics applications, an expansion into the third dimension with magnetic Hopfions is very attractive and may allow the discovery of novel physical phenomena. Since Hopfions are truly three-dimensional in nature, and therefore there is no two-dimensional approximation, advances in nanoscale three-dimensional synthesis and specifically characterization are essential to make progress.

Theoretical models and simulations predict that Hopfions can be stabilized in a variety of magnetic systems, including frustrated magnets[27], and, most promisingly, chiral magnets hosting target skyrmions (TSks)[3]. Target skyrmions[16,28,29] are extended skyrmionic spin textures, where the magnetization rotates multiple times (nπ)[30] and which have a topological charge that toggles from 0 to 1 for even and odd number of rotations n, respectively. They have been found in nanostructures of magnetic multilayers with strong DMI and perpendicular magnetic anisotropy (PMA), and are expected to host stable Hopfions by tuning the PMA at their interfaces. In particular, a 2π TSk transforms into a Hopfion when the magnetization at the top and bottom of the TSk changes to align with the magnetization at the center of the TSk. Fig 1b) displays 3D schematics of the spin textures indicating characteristic differences between the TSk and the Hopfion near the top and bottom of the structures.

To experimentally validate those theoretical predictions, we fabricated via magnetron sputtering two magnetic multilayer systems which were designed to host either a TSk or a magnetic Hopfion and will be in the following referred to as S7 (TSk) and S30 (Hopfion), resp. (Fig 1a). The multilayer structure of S7 consists of {Ir(1nm)/Co(1.5nm)/Pt(1nm)}x7, i.e., 7 layers of a PMA stack, and similarly, S30 consists of

{Ir(1nm)/Co(1nm)/Pt(1nm)]}x10 + {Ir(1nm)/Co(1.5nm)/Pt(1nm)}x10 + {Ir(1nm)/Co(1nm)/Pt(1nm)}x 10, i.e., 30 layers of a PMA stack with a varying Co thickness (Fig 1a). As a result of the interface induced PMA by Ir and Pt into Co, which is larger for the thinner Co layer, the PMA value in S7 with the 1.5nm thick Co layer is reduced to about 25-30% compared to S30 with a 1nm thick Co layer. To confine locally the spin textures and to harness also shape anisotropy, which supports the formation of TSks and Hopfions, nanodisks with a thickness of 24.5nm (S7) and 95nm (S30), respectively, and with varying diameters ranging from 2000nm down to 100nm were fabricated using electron beam lithography. The domain structure of S7 in a disk with a diameter of 1500nm as seen with X-PEEM is shown in Fig 1 of the Supplementary materials , and is found to be similar to the TSk textures observed previously with MTXM[16].

To distinguish the characteristic three-dimensional nature of the spin textures in TSks and Hopfions (Fig 1b), we combined two advanced magnetic X-ray microscopy techniques each with about 20-30nm lateral resolution, and both using X-ray magnetic circular dichroism (XMCD) as element-specific magnetic contrast mechanism. X-ray photoemission electron microscopy (X-PEEM) provides information about the spin texture close to the surface due to the limited escape depth of the detected secondary photoelectrons, while magnetic soft x-ray transmission microscopy (MTXM) images the local spin texture averaged throughout the thickness of the system. Further, XMCD measures the projection of the magnetization onto the photon propagation directions, which not only allows distinguishing in-plane vs out-of-plane components of the local magnetization, but also a reconstruction of the 3D spin texture by recording images at various incidence angles of the X-rays. Fig 1c shows simulations of experimental X-PEEM and MTXM images for a TSk and a Hopfion at a specific incidence angle, respectively. The MTXM images for the TSk and Hopfion exhibit both a pronounced black area at the center ("bulls-eye"), surrounded by a uniform bright ring and again a smaller black ring at the perimeter indicating that for a Q=1 Hopfion and 2π TSk MTXM images cannot distinguish between those two spin textures. However, the simulated XPEEM images (Fig 1c left column) show already at a single incidence angle distinct differences between the two spin textures. The TSk (Fig 1c top) exhibits a similar ring pattern to the MTXM image, whereas the Hopfion PEEM image exhibit a "ying-yang"-like pattern of a black and white structure each filling about half of the disk

Our experimental results obtained with the 400nm diameter pillars of the S30 sample are shown in Fig 2. The X-PEEM data were taken at PEEM3 (BL 11.0.1.1)[31] at the Advanced Light Source in Berkeley, CA, and the MTXM images shown were recorded at the MISTRAL beamline (BL-09)[32] at the ALBA synchrotron in Barcelona/Spain. Both X-PEEM and MTXM data were recorded at remanence, after the samples had been saturated by an external magnetic field. This approach is consistent with previous studies on TSks[16]. The MTXM images were recorded with the X-ray propagation direction orthogonal to the sample surface to maximize the XMCD contrast from out-of-plane magnetization. The X-PEEM images were recorded at an incidence angle of 30° with respect to the sample surface. To obtain a full 3D reconstruction of the spin textures in the surface region, X-PEEM images were recorded with the sample rotated around an axis normal to the surface by 0°, 90°, and 180°.

Fig 2a shows the simulation of the full 3D spin texture as it is expected in the X-PEEM image. The experimental X-PEEM data shown in Fig 2 b show a characteristic black-white feature as expected from the simulation at a single incidence angle (Fig 1c). The corresponding X-PEEM at a rotation of 90° shows the expected rotation of this black-white pattern. Combining the X-PEEM images from all three incidence angles retrieves the reconstructed 3D rendering of S30 shown in Fig 2c, which is in excellent agreement with the expected simulated spin texture (Fig 2a) taking into account the limited spatial resolution of X-PEEM. In the SI we present simulations of other 3D spin textures, that could exist in such multilayers, notably a magnetic toron[33] or a TSk. The distinct characteristics of the X-PEEM 3D spin texture for a magnetic Hopfion is a strong and unambiguous proof of the magnetic Hopfion textures in our system.

The experimental MTXM images recorded at the exact same disk as the X-PEEM images is shown in the bottom row of Fig 2. The characteristic bulls-eye structure (Fig 2c), which consists of an out-of-plane magnetization locally confined in the center of the disk and surrounded by a smooth magnetization pointing into the opposite directions is clearly visible. Even a faint black ring at the perimeter seems to be resolved as seen in Fig 2f, although this could also be a result of interference fringes that are common in MTXM at the boundaries of well defined structures. In conclusion, although the MTXM alone cannot distinguish between a TSk and a Hopfion, the experimental observation is compatible with the simulation, and the full 3D reconstruction of the X-PEEM data with its characteristic spin structure provides extreme strong evidence to the existence of a magnetic Hopfion in addition to the theoretical predictions in those magnetic multilayers.

To study in detail the variation of the spin textures as a function of PMA, further theoretical simulations are presented in Fig 3, which show that sandwiching a TSk hosting layer between two layers with high PMA will stabilize a Hopfion for all PMA values as long as the out of plane components of the TSk does not reach the surface of the magnetic material. With decreasing PMA, the Hopfion will extend further into the top and bottom PMA layers until finally a target skyrmion forms. These theoretical predictions are consistent with our experimental X-PEEM observations in S30 showing that the PMA domain structure does not extend throughout the magnetic multilayer system.

In conclusion, we have presented experimental evidence of the stabilization of magnetic Hopfions that were created in magnetic Ir/Co/Pt multilayers with tailoring their PMA values, and shaped into nanoscale disks. Using a combination of surface-sensitive X-ray photoemission electron microscopy and bulk-sensitive soft x-ray transmission microscopy we have observed in 400nm diameter disks characteristic features of a Hopfion in excellent agreement with predictions from theory. The out-of-plane bulls-eye texture located at the center of the nanodisk does not extend through the whole thickness of the sample, and a 3D rendering of the magnetic contrast at the surface matches that of a Q=1 Hopfion which is stabilized by interfacial DMI.

Our results, which mark an experimental milestone in the search for stable magnetic Hopfions, should not only stimulate further experimental and theoretical studies to search for Hopfions in magnetic nanostructures with different topologies and Hopf numbers, but should also open the door to exploiting novel phenomena in magnetic Hopfions, including Hopfion dynamics[34,35] that could lead to new applications in three-dimensional spintronic applications[36,37]. For example, compared to magnetic Skyrmions, the vanishing gyrovector in Hopfions would enable racetrack devices, which are not impacted by undesirable Hall effects[34].

## Methods

### Sample Fabrication

The design of the samples took into consideration various specific requirements of the x-ray spectromicroscopy techniques that were primarily used for their characterization, i.e. Magnetic Transmission Soft X-ray Microscopy (MTXM) and X-ray Photoemission Electron Microscopy (X-PEEM). To provide sufficient transmission of the soft x-rays between 700—800eV photon energy, commercially available 100nm thin amorphous Silicon Nitride (a-SiN$_x$) membranes were chosen as substrates. They exhibit an 85% transmission at the Co L$_3$ edge at 778eV, where most of the measurements were taken. To avoid potential electrostatic interference with the edges of the sample holder in the X-PEEM instrument large 1cm x 1cm Si frames were chosen with a 3mm x 3mm Si$_3$N$_4$ membrane window in the center.

The magnetic multilayers were patterned into nanopillars of varying diameters ranging from 2000nm to 100nm using Electron Beam Lithography (EBL) at the Molecular Foundry in Berkeley CA. Polymethyl methacrylate (PMMA) resist with aquaSAVE™, a conductive polymer used to alleviate charging problems, was exposed with a Vistec VB300 EBL tool and then developed in a 3:1 ratio of isopropyl alcohol: methyl isobutyl ketone (IPA:MIBK).

After development, magnetic multilayers were deposited via DC magnetron sputtering using a five-gun AJA ATC2000 confocal magnetron sputtering system with a liquid N$_2$ cryojacket achieving a base pressure <8x10$^{-8}$ torr. Layers were deposited with an Argon pressure of 1.3mtorr and at a rate of 0.04 nm/s. The roughness of the multilayers was measured with Atomic Force Microscopy (AFM) showing a smooth surface with an RMS <0.1nm.

After the deposition of the magnetic multilayers on the developed PMMA, an extended liftoff in dichloromethane was done, with the membranes upside down to alleviate strain. A 2nm Pt overlayer was deposited after liftoff to alleviate charging problems due to the low electrical conductivity of a-SiN$_x$ with X-PEEM.

To characterize the macroscopic magnetic properties of the magnetic multilayers, M(H) hysteresis curves were measured using a Lakeshore model 7400 vibrating sample magnetometer (VSM). They matched well with Ir/Co/Pt multilayers that host TSk[16], and with published work by others[12]. To confirm the increase in PMA for the Co 1nm multilayers a reference system {Ir(1nm)/Co(1.5nm)/Pt(1nm)}x10, {Ir(1nm)/Co(1nm)/Pt(1nm)}x10 multilayers was used (not shown in this paper). A first order approximation of PMA, i.e., for $K_u$ using $H_k = 2K_u/M_s$, yields a value for $K_u$ in the 1nm Co stack that is 3-4 times larger than in the 1.5nm Co stack which is in agreement with simulations where Hopfions are stabilized (Fig 3).

**Magnetic soft X-ray microscopy**

To characterize the distinct three-dimensional spin texture in TSks and Hopfions, two complementary advanced soft x-ray microscopy techniques were used, X-ray PhotoElectron Emission Microscopy (X-PEEM) and Magnetic Transmission X-ray Microscopy (MTXM). Both use X-ray magnetic circular dichroism (XMCD) as element-specific magnetic contrast mechanism. Large XMCD effects occur in the vicinity of element-specific inner core x-ray absorption edges, e.g. at the L-edges of transition metal such as Fe, Co, Ni or the M edges of rare earth systems. XMCD measures the dependence of the x-ray absorption coefficient of circularly polarized x-rays relative to the direction of the magnetization of a ferromagnetic material and scales with the projection of the magnetization onto the photon propagation direction of the X-rays. In combination with a laterally resolved detection of this XMCD contrast, images of the magnetic domain structures can be recorded in magnetic x-ray microscopies.

In X-PEEM the incoming monochromatic X-ray beam illuminates the specimen at a shallow angle, which for the PEEM3 system at the Advanced Light Source in Berkeley CA, where the data for this research were taken, amounts to 30° to the surface of the sample. Therefore, X-PEEM has an increased sensitivity to the in-plane component, although also the out-of-plane component can be detected. A rotating sample holder allows for in-situ rotation of the sample to determine the full 3D spin texture near the surface. X-PEEM detects the secondary photoelectrons generated in the x-ray absorption process, which leave the surface and are then transmitted through an electronic optical system onto at 2D CCD devices, which records the emitted electrons. Since the escape depth of those secondary electron is limited to about 5nm, X-PEEM images the spin texture from that depth only. With advanced aberration-corrected PEEM systems, a spatial resolution down to about 20nm has been reported[38]. Magnetic X-PEEM experiments are generally performed at remanence to avoid serious distortion of the electron beams by varying external magnetic fields.

The MTXM system is built in analogy to an optical microscopy consisting of the source, a condenser optics, a high-resolution objective lens, which for x-rays is a Fresnel zone plate (FZP), and a two-dimensional back-illuminated CCD detector. At the MISTRAL beamline[32], where the MTXM images were recorded, the optical system consists of a PGM monochromator which provides monochromatic x-rays. A capillary condenser system collimates the photon and provides a uniform illumination of the specimen. An image of the transmitted photons is formed by the downstream high-resolution micro-zone-plate (MZP) of 25nm outermost zone width and directly recorded by a CCD device. The spatial resolution in an MTXM is primarily determined by the quality of the MZP and has been shown to be able to less than 10nm[39]. The MZP at MISTRAL provides about 20-30nm spatial resolution. MTXM detects directly the transmitted photons and therefore, if the incident photon intensity is known, the actual x-ray absorption coefficient can be derived following Beer's Law. In normal incidence, the XMCD contrast in MTXM detects the out-of-plane magnetic component only. In order to image in-plane components, the sample is typically rotated around an axis perpendicular to the x-ray propagation direction. In general, MTXM can image magnetic domains in varying external magnetic fields, however, for this study all images were recorded in remanence.

To allow for a direct comparison of the X-PEEM and MTXM images, the same sample, which was fabricated on x-ray transparent a-SiN$_x$ membranes, was used for both the X-PEEM and the MTXM experiments, to ensure that the identical structure was imaged with both techniques. Both X-PEEM and MTXM are full-field x-ray microscopy techniques covering a field-of-view of several µm in a single image. Typical image acquisition times were a few seconds per image for both MTXM and X-PEEM. The spectral resolution for both X-PEEM and MTXM was sufficiently high at about ΔE/E ~ 10$^3$. All measurements were performed at room temperature.

X-PEEM and MTXM images were recorded as 2dim arrays of 1024x1024 px CCD data.

**Data analysis**

The experiments shown in this paper were recorded at the Co L$_3$ absorption edge. The MTXM data were taken at fixed circular polarization, and standard image processing techniques were applied to take into account the much larger count rate in the areas surrounding the disks due to the much lower x-ray absorption than in the disk itself.

To separate the in-plane from the out-of-plane components in the X-PEEM images, four series of images were recorded at each location and for each incident angle θ: left circularly polarized x-rays at the resonant Co L$_3$ edge (L_ON), 2) left circularly polarized light below the Co resonance at 770 EV (L_OFF) and vice versa for right circularly polarized x-rays (R_ON, and R_OFF, respectively).

Normalized XMCD images were derived according to Image(θ)= (L_ON)/(L_OFF) - (R_ON/R_OFF) which allows for quantitative comparison of the magnetization when imaged at different angles. Since this normalization process removes the non-magnetic background in X-PEEM, the surrounding areas of the disks appear grey, whereas they appear white (=saturated) in the MTXM images.

The variation of incident angles in X-PEEM, i.e., images measured at 0°, 90°, and 180°, where 0° is chosen as an arbitrary value allows to extract the out-of-plane component: Image(OOP) = Image(0°) + Image(180°), since the in-plane component points in opposite direction and therefore cancels out. Similarly, the combinations Image(0°) – Image(OOP)/2 and Image (90°) - Image(OOP)/2 provides the in-plane components. The combination of in-plane and out-of-plane components was used to finally retrieve the full 3D orientation of the spin textures in the X-PEEM images.

Imaging isolated nanostructures in MTXM and XPEEM can create artefacts at the boundaries originating from different sources. In XPEEM charge can build up near the edge of isolated structures, resulting in a reduced area of magnetic contrast compared to the physical nanostructure. In MTXM, unless the XMCD can be precisely and without image distortion modulated by polarization reversal, the abrupt change in transmission at the edge of the nanostructure can lead to interference fringes that are challenging to remove through filtering.

**Theoretical simulations**

The simulations were performed to mimic the anticipated transformation of TSks into Hopfions by increasing the PMA at the top and bottom of the TSk. Effective medium models work well with simulating thin films because the magnetic properties of the material are relatively uniform over the thickness of the film. To simulate the S30 system, where this assumption of uniformity does not hold, the anisotropy was varied through the film thickness.

As experimentally measured spatial profiles of the PMA of each individual Co layer are not available for our system, we approximated it by an assembly of three magnetic multilayers with the same DMI and the same thickness. The PMA in each of the three subsystems was assigned a spatially varying PMA, e.g., as a trial function the linearly varying PMA distribution ($K = \frac{3}{52} + \frac{6}{13}x$), where x is the distance from the middle of the central layer, measured in units of the thickness of a layer and PMA being dimensionless. The top and bottom third was assigned the same PMA distribution but with a value that is higher than the PMA in the central third, e.g. 3:1. Given that the actual variation of PMA in S30 is not known, a variety of simulations were performed with multiple profiles of varying the PMA between the central layer and the top/bottom layer.

One set of simulations followed a model published previously[3] where the central layer has no PMA and the top and bottom layer have a uniform PMA. It was found that above a certain critical value of PMA the out-of-plane structure of TSks no longer reached the top or bottom layer of the material, but a Hopfion was formed. Below this critical PMA value, only TSks were found to be stable. While this approach helps account for the effects of a reduced PMA due to shape anisotropy, it is likely that the central layer still has a non-negligible effective PMA, and that the PMA in all layers varies spatially over each layer.

Another set of simulations implemented a linearly varying PMA in all layers, with the PMA value being smallest in the central layer and increasing outwards, allowing for a relative PMA ratio between the different PMA layers to be quantitatively described. It was found that when the PMA ratio of the high PMA layer to the low PMA layer is 3:1 or greater a Hopfion can be stabilized when the out of plane structure of the TSk does not extend through the thickness of the material (Fig 3). Below this ratio, Hopfions are not stabilized. This is consistent with VSM measurements of {Ir(1nm)/Co(1nm)/Pt(1nm)}x10 multilayers compared to {Ir(1nm)/Co(1.5nm)/Pt(1nm)}x10 multilayers showing a PMA ratio of 3-4 between the high (Co=1nm) and low (Co=1.5nm) Co film.

None of these simulations included the demagnetizing field (stray field energies) which is an acceptable approximation for several reasons. Compared to TSks, Hopfions generate much weaker stray fields[4], which means that the addition of stray field energies *increases* the likelihood of a Hopfion being stabilized energetically. Further, the additional in-plane shape anisotropy generated from the disk structure as a result of the demagnetizing field is accounted for in the simulations by setting the PMA value of the low PMA layers to zero.

The simulated MTXM and X-PEEM patterns shown in Fig. 1c) were derived from these simulations using the actual wave vector for the incoming x-rays, i.e., perpendicular to the sample surface for MTXM and therefore only sensitive to the out-of-plane component of magnetization, and at an inclined angle of 30° for X-PEEM which is sensitive to both in plane and out of plane components of magnetization

**Acknowledgements**

We would like to thank Robert Streubel for stimulating discussions.

This work was funded by the U.S. Department of Energy, Office of Science, Office of Basic Energy Sciences, Materials Sciences and Engineering Division under Contract No. DE-AC02-05-CH11231 (Non-equilibrium magnetic materials program MSMAG).

Work at the Molecular Foundry was supported by the Office of Science, Office of Basic Energy Sciences, of the U.S. Department of Energy under Contract No. DE-AC02-05CH11231.

This research used resources of the Advanced Light Source, which is a DOE Office of Science User Facility under contract no. DE-AC02-05CH11231.

This research includes experiments that were performed at MISTRAL beamline at ALBA Synchrotron in collaboration with ALBA staff.

AH-R. acknowledges the support from European Union's Horizon 2020 research and innovation program under Marie Skłodowska-Curie grant ref. H2020-MSCA-IF-2016-746958.

**Author Contributions**

NK and PF conceived the conception and design of the experiment with guidance from the theoretical input by PS. NR supported the design, fabricated the multilayers, and performed the VSM measurements. NK, DR, IC, SV, and SD performed the nanofabrication tasks at the Molecular Foundry. NK, PF, AH-R., AS, EP, and SF performed the MTXM experiments at ALBA. NK, DR, and RC performed the X-PEEM experiments at the ALS. NK and DR analyzed the data. PS provided the theoretical simulations with input from NK. NK and PF wrote the manuscript. All authors provided critical input to the scientific discussion, and proofreading the manuscript.

**Competing interests**

The authors declare no competing interests.

**Materials & Correspondence**

All requests should be directed to Peter Fischer *PJFischer@lbl.gov*, or Noah Kent *Nakent@ucsc.edu*

**Figures**

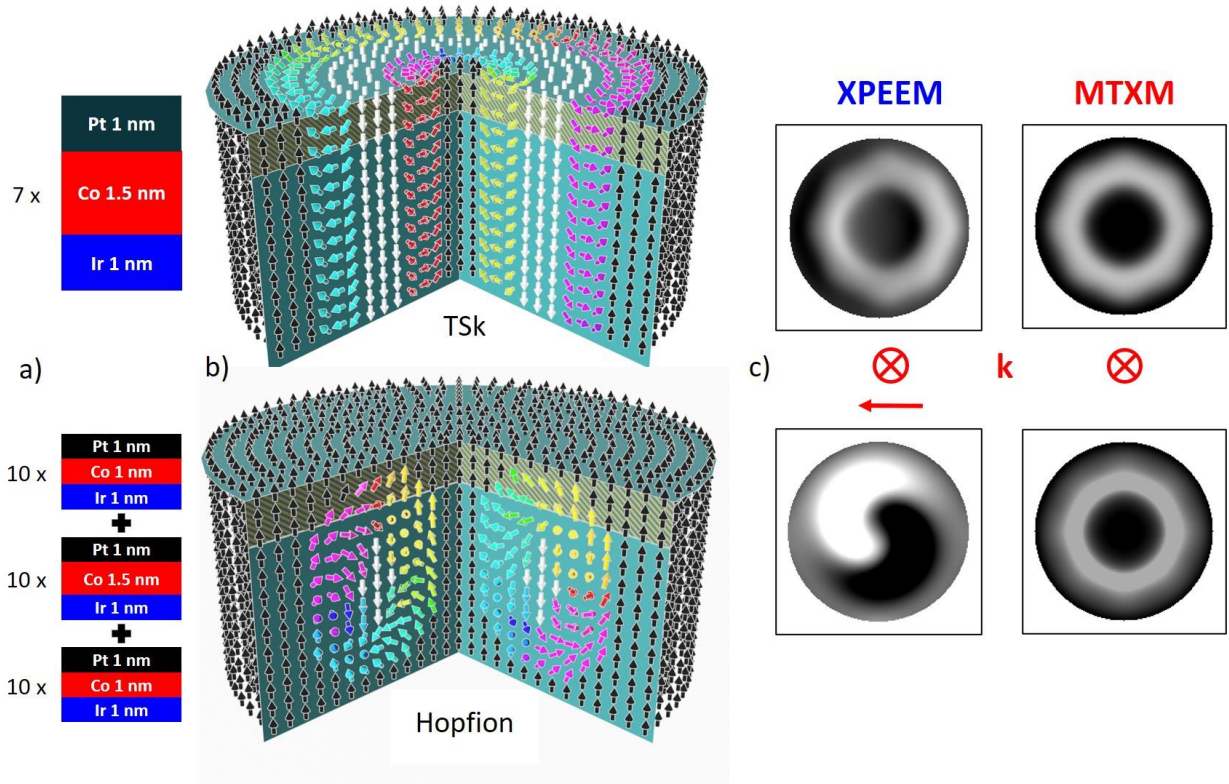

**Figure 1. Spin texture of a 2π TSk and a Q=1 Hopfion. a)** Schematics of the Ir/Co/Pt multilayer systems (S7, S30) used in this study. b) Schematical drawing of the spin texture of a 2π TSk (top) and a symmetric Q=1 Hopfion (bottom) . The yellow shaded region near the top indicates the approximate depth sensitivity of X-PEEM. **c)** Simulated X-PEEM (left) and MTXM (right) signals for a TSk (top) and a Hopfion (bottom) The direction of the photon angular momentum (k) is indicated for X-PEEM (in-plane and out-of-plane) and MTXM (only out-of-plane).

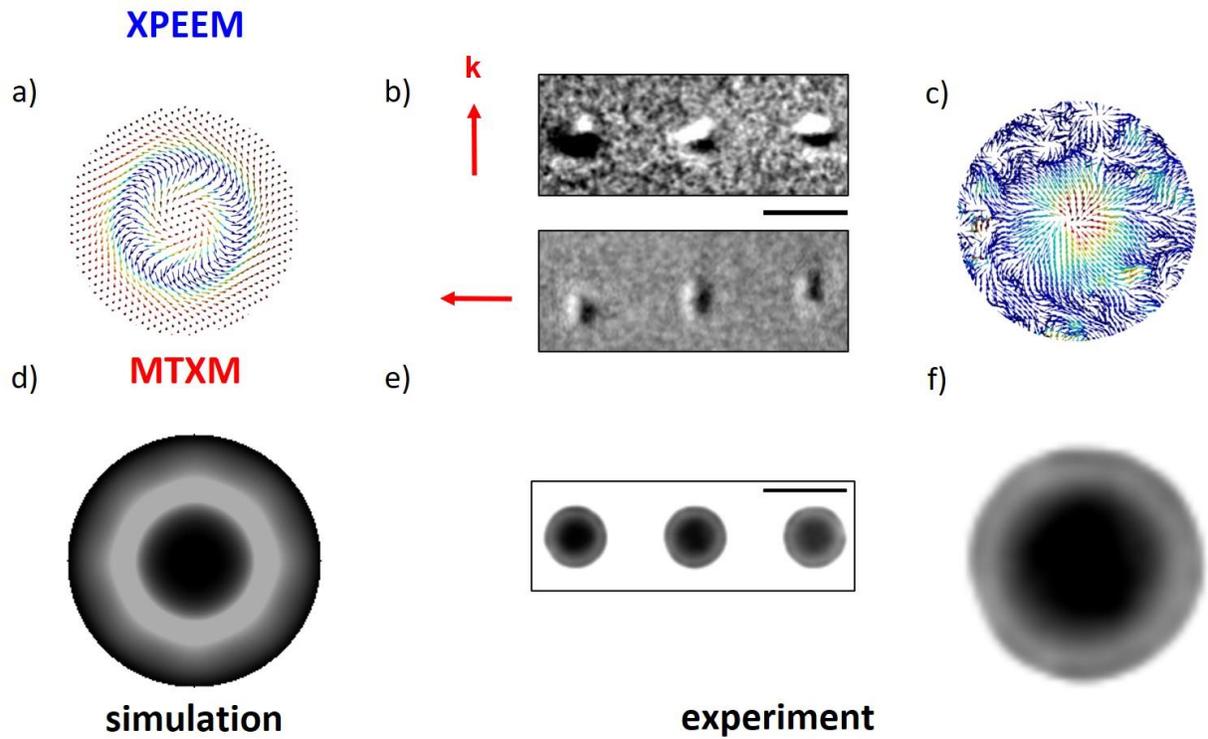

**Figure 2. Simulation and Experimental results for X-PEEM and MTXM a)** Simulation of the full 3D spin textures expected in X-PEEM. **b)** Rows of three S30 nanodisks with 400nm diameter observed with XPEEM under θ=0° and θ=90° **c)** 3D spin texture reconstructed from experimental X-PEEM images. **d)** Simulation of the MTXM image. **e)** MTXM images of a row of three S30 nanodisks as in b) showing a reproducible characteristic spin texture. **f)** MTXM image of the identical disk as in c). Scale bars in b) and e) are 500nm.

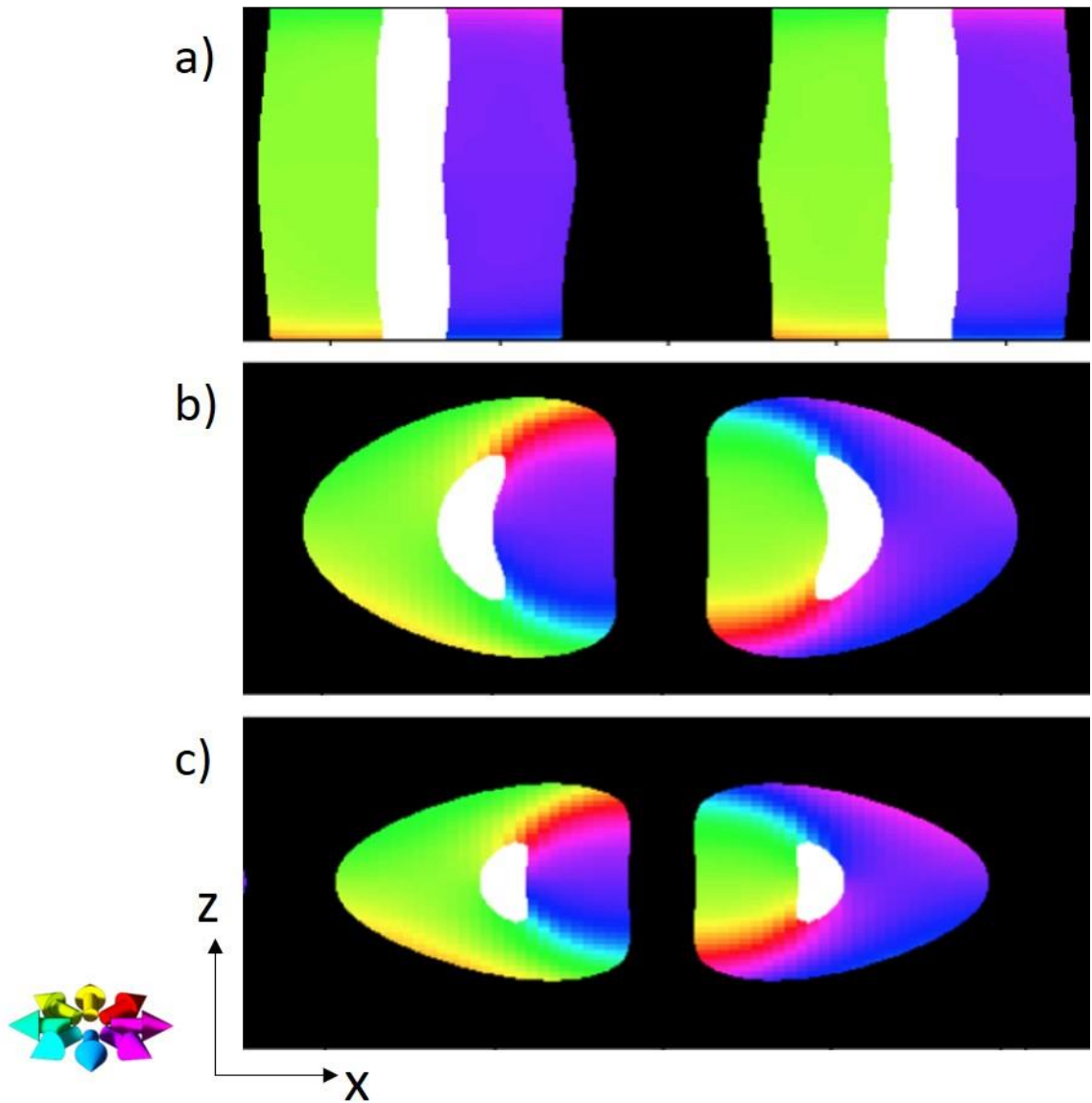

**Figure 3. Theoretical simulations.** A slice along the x-z plane (z along the disk height) of a variable PMA magnet with DMI. Color code as in Fig 1b, i.e. black corresponds to up, and white to down, and the color wheel refers to the in plane (x-y plane) magnetization direction. The PMA in the central third of each simulated layer is lower than the PMA of the top, bottom third of each simulation. The ratio of high to low PMA is a) 2:1, b) 3:1, c) 4:1.

# SUPPLEMENTARY INFORMATION

## Creation and confirmation of Hopfions in magnetic multilayer systems

Noah Kent[1,2], Neal Reynolds[1,3], David Raftrey[1,2], Ian T.G. Campbell[1,3], Selven Virasawmy[4], Scott Dhuey[4], Rajesh V. Chopdekar[5], Aurelio Hierro-Rodriguez[6], Andrea Sorrentino[7], Eva Pereiro[7], Salvador Ferrer[7], Frances Hellman[1,3], Paul Sutcliffe[8], Peter Fischer[1,2]

[1]Materials Sciences Division, Lawrence Berkeley National Laboratory, Berkeley, CA 94720, USA
[2]Physics Department, UC Santa Cruz, Santa Cruz CA 95064, USA
[3]Department of Physics, University of California, Berkeley, Berkeley, CA 94720, USA
[4]The Molecular Foundry, Lawrence Berkeley National Laboratory, Berkeley, CA 94720, USA
[5]Advanced Light Source, Lawrence Berkeley National Laboratory, Berkeley, CA 94720, USA
[6]Department of Physics, University of Oviedo, 33007 Oviedo, Spain
[7]ALBA Synchrotron, 08290 Cerdanyola del Vallès, Spain
[8]Department of Mathematical Sciences, Durham University, Durham DH1 3LE, UK

The supplementary material in this section will provide further support to the confirmation of a magnetic Hopfion in the two multilayer systems studied.

- Experimental X-PEEM data confirm the TSk spin texture in the sample S7.
- Extended simulations of torons as another possible 3D spin texture in those systems show a disagreement with the experimental X-PEEM images.

## S1. X-PEEM images of the S7 multilayer

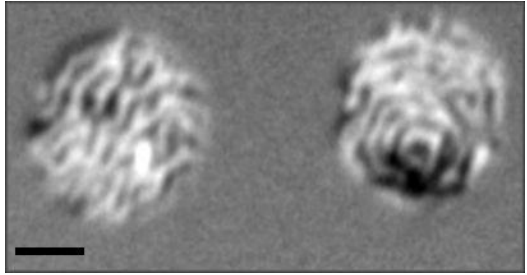

The domain structure of the S7 multilayer system in a disk with a diameter of 1500nm imaged with X-PEEM at the Co $L_3$ absorption edge is found to be similar to the TSk textures observed previously with MTXM[1]. Notably, at the bottom portion of the right disk shown above a $2\pi$ TSk is clearly visible. We conclude that S7 is hosting TSks with a characteristic PMA domain structure that extends through the thickness of the nanodisk. The scale bar shown is 500nm.

**S2. Extended micromagnetic simulations in magnetic multilayers.**

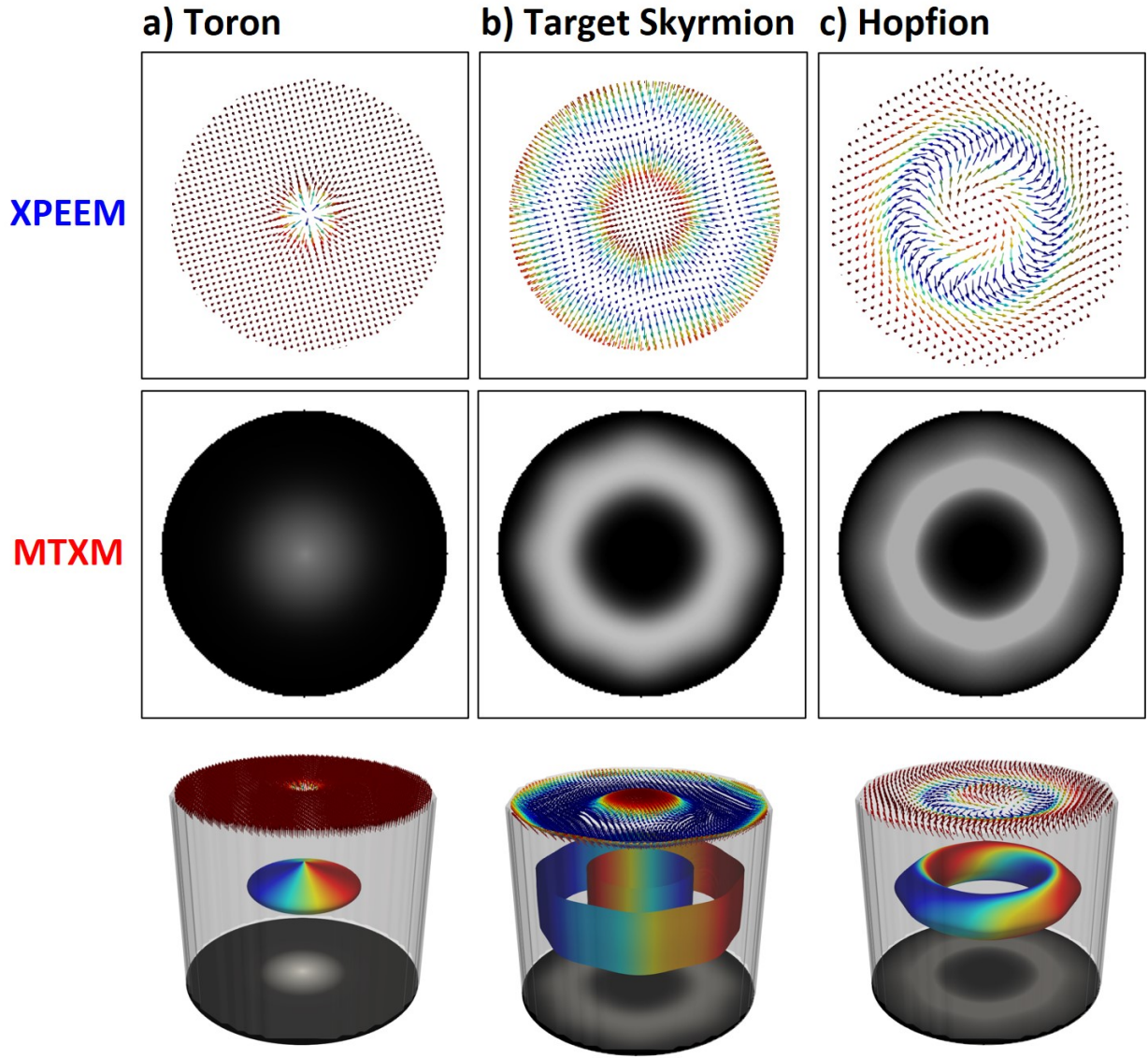

To exclude other possible 3D spin textures in those magnetic multilayers, e.g. magnetic torons[2], we performed micromagnetic simulations to determine the expected X-PEEM and MTXM images in those structures. The results are summarized in the figure above with the X-PEEM simulations in the top row and, the MTXM simulations in the central row. The bottom row is an artistic drawing, combining the simulated X-PEEM and MTXM images at the top and the bottom of a cylinder, and the 3D winding of the spin textures inside the disk positioned in the center of the cylinder.

a) The magnetic toron shows in the X-PEEM image - apart from a radially symmetric area in the center - a rather uniform magnetization pointing into the opposite direction than in the center. Similarly, the

MTXM image is expected to show a relatively small light center surrounded by a uniform magnetization, pointing into the opposite directions (black).

b) The target skyrmion shows the expected characteristic multi-ring structure, which for the MTXM image exhibits a black central area followed by a bright ring and then a wide, black ring at the perimeter. The X-PEEM image shows a similar multi-ring structure as the MTXM, indicative of the multiple rotations of the magnetization into radial outward direction of the disk.

c) The Hopfion structure, although quite similar to the TSk for the MTXM image, exhibits a X-PEEM image that is in stark difference to both the toron and the TSk.

Comparing those three spin textures with the experimental data, c.f. Figs 2c and 2f in the main manuscript, confirms that our experimental data are in agreement with the Hopfion texture.